\documentclass[12pt,preprint,epsf]{aastex}
\usepackage{natbib}

\slugcomment{ }
\shorttitle{SN 2014j}
\shortauthors{ et al.}
\usepackage{epstopdf}
\usepackage{graphicx}
\usepackage{amsmath}

\newcommand\numberthis{\addtocounter{equation}{1}\tag{\theequation}}

\begin{document}
\title{Late-time Flattening of Type Ia Supernova Light Curves: Constraints From SN\,2014J in M82}
\shorttitle{Late-time SN\,2014J}
\author{Yi Yang\altaffilmark{1,2}, 
        Lifan Wang\altaffilmark{1,3},
        Dietrich Baade\altaffilmark{4}, 
        Peter.~J. Brown\altaffilmark{1},
        Aleksandar Cikota\altaffilmark{4}, 
        Misty Cracraft\altaffilmark{5}, 
        Peter A. H\"oflich\altaffilmark{6}, 
        Justyn R. Maund\altaffilmark{7}{$^{,\dagger}$}, 
        Ferdinando Patat\altaffilmark{4},
        William~B. Sparks\altaffilmark{5}, 
        Jason Spyromilio\altaffilmark{4},
        Heloise F. Stevance\altaffilmark{7}, 
        Xiaofeng Wang\altaffilmark{8}, 
        J. Craig Wheeler\altaffilmark{9}}

\altaffiltext{1}{George P. and Cynthia Woods Mitchell Institute for 
Fundamental Physics $\&$ Astronomy, Texas A. $\&$ M. University, 
Department of Physics and Astronomy, 4242 TAMU, College Station,
TX 77843, USA, email: yi.yang@weizmann.ac.il}
\altaffiltext{2}{Department of Particle Physics and Astrophysics, 
Weizmann Institute of Science, Rehovot 76100, Israel}
\altaffiltext{3}{Purple Mountain Observatory, Chinese Academy 
of Sciences, Nanjing 210008, China}
\altaffiltext{4}{European Organisation for Astronomical Research 
in the Southern Hemisphere (ESO), Karl-Schwarzschild-Str. 2, 85748 
Garching b.\ M{\"u}nchen, Germany}
\altaffiltext{5}{Space Telescope Science Institute, Baltimore, 
MD 21218, USA}
\altaffiltext{6}{Department of Physics, Florida State University, 
Tallahassee, Florida 32306-4350, USA}
\altaffiltext{7}{Department of Physics and Astronomy, University of 
Sheffield, Hicks Building, Hounsfield Road, Sheffield S3 7RH, UK}
\altaffiltext{8}{Physics Department and Tsinghua Center for 
Astrophysics (THCA), Tsinghua University, Beijing, 100084, China}
\altaffiltext{9}{Department of Astronomy and McDonald Observatory, 
The University of Texas at Austin, Austin, TX 78712, USA}
\altaffiltext{$\dagger$}{Royal Society Research Fellow}

\begin{abstract}
The very nearby Type Ia supernova 2014J in M82 offers a rare 
opportunity to study the physics of thermonuclear supernovae 
at extremely late phases ($\gtrsim$800 days). Using the 
{\it Hubble Space Telescope}, we obtained six epochs of high 
precision photometry for SN\,2014J from 277 days to 1181 days 
past the $B-$band maximum light. The reprocessing of electrons 
and X-rays emitted by the radioactive decay chain 
$^{57}$Co$\rightarrow ^{57}$Fe are needed to explain the 
significant flattening of both the $F606W$-band and the 
pseudo-bolometric light curves. The flattening confirms 
previous predictions that the late-time evolution 
of type Ia supernova luminosities requires additional energy 
input from the decay of $^{57}$Co \citep{Seitenzahl_etal_2009}. 
By assuming the $F606W$-band luminosity scales with the bolometric 
luminosity at $\sim$500 days after the $B-$band maximum light, 
a mass ratio $^{57}$Ni/$^{56}$Ni$\sim$0.065$_{-0.004}^{+0.005}$ 
is required. This mass ratio is 
roughly $\sim$3 times the solar ratio and favors a progenitor 
white dwarf with a mass near the Chandrasekhar limit. 
A similar fit using the constructed pseudo-bolometric 
luminosity gives a mass ratio 
$^{57}$Ni/$^{56}$Ni$\sim$0.066$_{-0.008}^{+0.009}$. 
Astrometric tests based on the multi-epoch {\it HST} ACS/WFC 
images reveal no significant circumstellar light echoes in 
between 0.3 pc and 100 pc \citep{Yang_etal_2017a} from the 
supernova. 
\end{abstract}

\section{Introduction \label{intro}}
The astronomical community widely agrees that luminous 
hydrogen-poor Type Ia supernovae (SNe) explosions are powered 
by the thermonuclear runaway of ($\geqslant 1 M_{\odot}$) 
carbon/oxygen white dwarfs (WDs \citealp{Hoyle_etal_1960}). 
The accretion-induced explosion fuses $\sim$0.1-1.0$M_{\odot}$ 
of radioactive $^{56}$Ni. {Type Ia SNe cosmology uses 
these SNe as the most accurate distance indicators at redshifts 
out to $z\sim$2 \citep{Riess_etal_1998, Perlmutter_etal_1999, Riess_etal_2016}}. 
Amazingly, this accuracy is achieved without knowing the exact  
nature of the progenitors. 

Prior to maximum luminosity, the light curve of Type Ia SNe is powered 
by the energy generated by the decay of explosion-synthesized 
radioactive nuclei. The reprocessing in the ejecta 
converts the energy to longer wavelengths. The decay chain of 
$^{56}$Ni$\rightarrow ^{56}$Co$\rightarrow ^{56}$Fe provides 
the main source of energy deposition into the ejecta of Type I 
SNe \citep{Arnett_etal_1982}. During the early phases, the 
optically-thick ejecta trap the energy. The dominant process 
is Compton scattering of $\gamma$-rays produced by the decay 
$^{56}$Ni + $e^{-} \rightarrow ^{56}$Co $+ \ \gamma \ + \ \nu_{e}$,  ($t_{1/2} \sim$6.08 days), which 
allows energy to escape as X-ray continuum or absorbed by the 
material in the ejecta via the photoelectric effect (see 
\citealp{Milne_etal_1999, Penney_etal_2014} for comprehensive reviews). 
The produced $^{56}$Co decays to stable $^{56}$Fe, and the $^{56}$Co decay process, 
with half-life $t_{1/2} \sim$77 days, dominates after $\sim$200 days, when the 
expanding ejecta become more and more optically thin, and the 
column density decreases as $t^{-2}$ (e.g., 
\citealp{Arnett_etal_1979, Chan_etal_1993, Cappellaro_etal_1997, 
Milne_etal_1999}). {Eighty-one percent of the 
$^{56}$Co decays via electron capture 
($^{56}$Co $+ \ e^{-} \rightarrow ^{56}$Fe $+ \ \gamma \ + \ \nu_{e}$), 
and the remainder decays through annihilation of high 
energy positrons in the ejecta 
($^{56}$Co $\rightarrow ^{56}$Fe $+ \ e^{+} \ + \gamma \ + \nu_{e}$). 
}

Observations at extremely late phases provide unique opportunities 
to examine various models exploring the effects of a magnetic 
field. As long as energy deposition is dominated by positrons 
being completely trapped by the magnetic field, the slope of the 
bolometric light curve should match the $^{56}$Co decay rate. 
{On the other hand, \citet{Milne_etal_1999} suggested  
a ``radially combed'' magnetic field, {or even a magnetic-field-free 
situation (as no magnetic field in radial directions will 
lead to an increasing fraction of positron escape), would cause the 
light curve to decline faster than the rate of $^{56}$Co decay.} 
The discrepancy between the ``trapping scenario'' with a confining 
magnetic field and the case without magnetic field can 
be as significant as 2 magnitudes in the photometric light curves 
from 400 - 800 days (see Figure 9 of \citealp{Milne_etal_1999}). 
Similar variations of the late-time light curves have been found 
by \citet{Penney_etal_2014} based on measuring positron transport 
effects and their dependency on the magnetic field with late-time 
line profiles. 
As the SN envelope undergoes homologous expansion, the morphology of the 
magnetic field remains {but the Larmor radius increases linearly with time, 
such that the fraction of escaped photons would exhibit a 
time-dependence due to the variations of the magnetic field and the} 
light curve should decline faster than the rate of $^{56}$Co decay. 

Additonally, different effects of nucleosynthesis can be testable 
through the very late photometric evolution of Type Ia SNe and may 
be used to discriminate between different explosion models. Two of 
the most favorable explosion channels: a delayed detonation in a 
Chandrasekhar-mass white dwarf \citep{Khokhlov_etal_1991} 
and a violent merger of two carbon-oxygen white dwarfs 
\citep{Pakmor_etal_2011, Pakmor_etal_2012}, will result in late-time 
light curves behaving differently due to different amounts of ejecta 
heating from $^{57}$Co and $^{55}$Fe \citep{Ropke_etal_2012}. 
{The decline rate of the light curve at 
extremely late times provides a unique opportunity, therefore, to test} 
the enigmatic explosion mechanisms of Type Ia SNe. 

Increasing evidence shows the flattening of Type Ia SN 
light curves around 800 to 1000 days, i.e., SN\,1992A ($\sim$950 days; 
\citealp{Cappellaro_etal_1997}, \citealp{Cappellaro_etal_1997}), 
SN\,2003hv ($\sim$700 days; \citealp{Leloudas_etal_2009}), and 
SN\,2011fe ($\sim$930 days; \citealp{Kerzendorf_etal_2014}). This 
flattening cannot be explained even by complete trapping of the 
$^{56}$Co positrons. \citet{Seitenzahl_etal_2009} suggested that 
additional heating from the Auger and internal conversion electrons, 
together with the associated X-ray cascade produced by the decay of 
$^{57}$Co$\rightarrow ^{57}$Fe ($t_{1/2}\approx$272 days) and 
$^{55}$Fe$\rightarrow ^{55}$Mn ($t_{1/2}\approx$1000 days), will 
significantly slow down the decline of the light curve. 

Only recently, \citet{Graur_etal_2016} carried out an analysis 
of the light curve of SN\,2012cg as late as $\sim$ 1055 days after 
the explosion and excluded the scenario in which the light curve 
of SN\,2012cg is solely powered by the radioactive decay chain 
$^{56}$Ni$\rightarrow ^{56}$Co$\rightarrow ^{56}$Fe, unless there is  
an unresolved light echo $\sim$14 magnitudes fainter 
than the SN peak luminosity. 
Another very careful study on the late-time evolution of SN\,2011fe has 
already extended the observing effort to an unprecedented 1622 days 
past the $B-$band maximum light \citep{Shappee_etal_2016}. This analysis 
has clearly detected the radioactive decay channel powered by $^{57}$Co, 
with a mass ratio of 
log($^{57}$Co/$^{56}$Co)$=-1.62^{+0.08}_{-0.09}$. This abundance ratio 
is strongly favored by double degenerate models which require a lower 
central density. The detection of $^{55}$Fe is still unclear at these late 
epochs \citep{Shappee_etal_2016}. 
Another study based on the pseudo-bolometric light curve 
for the SN\,2011fe has measured the mass ratio of $^{57}$Co to $^{56}$Co 
to be 1.3 -- 2.5 times the solar value, which is broadly consistent with the 
ratios predicted for the delayed detonation models \citep{Dimitriadis_etal_2017}. 
Additionally, spectroscopic information of the nearby 
SN\,2011fe has been obtained at 981 days \citep{Graham_etal_2015_11fe} 
and 1034 days \citep{Taubenberger_etal_2015}. Strong energy 
input from the radioactive decay of $^{57}$Co is required, without 
which the optical spectrum would be underproduced by a factor of $\sim$4 
\citep{Fransson_etal_2015}. The mass ratio of $^{57}$Ni to $^{56}$Ni 
produced, which gives a strong constraint on the Type Ia 
SN explosions, is found to be roughly 2.8 and 2 times of the solar ratio 
for SN\,2011fe and SN\,2012cg, respectively 
\citep{Fransson_etal_2015, Graur_etal_2016}. 

{Recently, \citet{Graur_etal_2017} proposed a new 
model-independent correlation between the stretch of a SN and the shape 
of their late-time light curves based on the shapes of the light curve of 
four type Ia SNe measured at $\textgreater$ day 900, i.e., 
SN\,2012cg \citep{Graur_etal_2016}, SN\,2011fe \citep{Shappee_etal_2016}, 
SN\,2014J (this work) and SN\,2015F \citep{Graur_etal_2017}. They indicated 
that $^{57}$Co may be underproduced in subluminous type Ia SNe.
This correlation provides a novel way to test various physical processes 
driving the slow-down of the type Ia SN light curves $\sim$900 days after 
explosion.
} 

SN\,2014J was first discovered on Jan 21.805 UT by \citet{Fossey_etal_2014} 
in the very nearby starburst galaxy M82 (3.53$\pm$0.04 Mpc, \citealp{Dalcanton_etal_2009}). 
Later observations constrained the first light of the SN to Jan.\ 14.75 UT
\citep{Zheng_etal_2014, Goobar_etal_2014}.  This date is consistent with 
the early rising recorded by the 0.5-m Antarctic Survey Telescope (AST) 
during its test observations \citep{Ma_etal_2014} as well as with other 
pre-discovery limits reported by various groups \citep{Denisenko_etal_2014, 
Itagaki_etal_2014, Gerke_etal_2014}. SN\,2014J reached its $B-$band maximum 
on Feb. 2.0 UT (JD 2,456,690.5) at a magnitude of 11.85$\pm$0.02 
\citep{Foley_etal_2014}. Follow-up photometric and spectroscopic 
observations have been made by various groups \citep{Lundqvist_etal_2015, 
Bonanos_etal_2016, Srivastav_etal_2016, Johansson_etal_2017}. 
The strength of $\gamma$-ray lines \citep{Churazov_etal_2014, Diehl_etal_2015} 
and an analytic model fit to the pseudo bolometric light curve 
\citep{Srivastav_etal_2016} of SN\,2014J suggest that $\sim$0.5-0.6 
$M_{\odot}$ of $^{56}$Ni {was} synthesized in the explosion. In this paper, 
we present our late time {{\it Hubble Space Telescope (HST)}} 
photometric observations of SN\,2014J and 
fit both the $F606W$ (broad $V$) band and an estimate of the pseudo-bolometric 
luminosity evolution with the Bateman equation considering the luminosity 
contributed by the decay of $^{56}$Co, $^{57}$Co, and $^{55}$Fe. In addition 
to following a similar approach presented in \citet{Graur_etal_2016}, we 
provide a careful astrometric analysis to the time-evolution of the position 
and profile of the SN\,2014J point source at very late epochs. 

\section{Observations and Data Reduction}
We imaged the SN\,2014J with the {\it Hubble Space Telescope} Advanced Camera 
for Surveys/Wide Field Channel ({\it HST} ACS/WFC) during {six visits (V1-V6)} 
under multiple {\it HST} programs: GO-13717 (PI: Wang), GO-14139 (PI: Wang), 
and GO-14663 (PI: Wang), i.e., V1$\sim$day 277, V2$\sim$day 416, V3$\sim$day 
649, V4$\sim$day 796, V5$\sim$day 983, and V6$\sim$day 1181 
relative to its $B-$band maximum at a mgnitude of 11.85$\pm$0.02 on 
Feb. 2.0 UT (JD 2,456,690.5, \citealp{Foley_etal_2014}). 
Figure~\ref{Fig_fov} shows the 
field around SN\,2014J. A log of observations is presented in Table \ref{Table_1}. 
Exposures obtained with different ACS visual polarizers and in different 
filter combinations and visits have been aligned through {\it Tweakreg} in 
the {\it Astrodrizzle} package \citep{Gonzaga_etal_2012}. 

\begin{figure}[!htb]
\epsscale{1.0}
\plotone{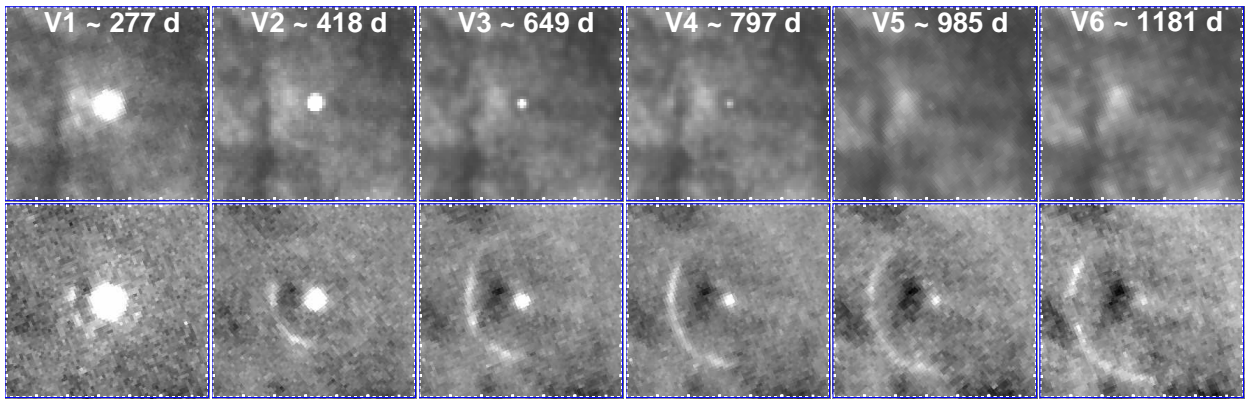}
\caption{\small {\it HST} ACS/WFC $F606W$ (upper panels) and associated 
$F606W-F555W$ (lower panels) images of SN\,2014J obtained in six 
different visits as labeled. Each square measures 3$\arcsec$.2 $=$ 54 pc 
along its sides (oriented such that north is up, east is left). The distance between little 
tick marks corresponds to 0$\arcsec$.1. 
{Resolved light echoes arising from 
interstellar dust clouds are observed at large foreground distances ($\gtrsim$100 pc) from 
the SN. A luminous arc is visible in the lower left quadrant and
a radially diffuse ring can be seen over a wide range in position angle. 
See \citet{Yang_etal_2017a} for more details.} 
\label{Fig_fov}} 
\end{figure}

{The throughput of each ACS/WFC polarizer being used by the 
{\it Synphot \footnote[1]{http://www.stsci.edu/institute/software\_hardware/stsdas/synphot}} 
synthetic photometry does not match the values determined from on-orbit calibrations.} 
We corrected the 
polarizers' throughput with the values deduced by on-orbit calibrations (i.e., 
Table 12 of \citealp{Cracraft_Sparks_2007}, also see \citealp{Biretta_etal_2004}). 
Following the three polarizers case described in earlier works by 
\citet{Sparks_Axon_1999}, we deduced the Stokes vectors from the observations. 
In this work, we only discuss the observed flux from SN\,2014J, and the 
intensity maps (Stokes I) are the only required input parameter for this analysis. 
\begin{align*}
I = \frac{2}{3} [r(POL0) + r(POL60) + r(POL120)],  
\numberthis \label{eqn_3}
\end{align*}
where $r(POL0)$, etc.\ are the count rates in the images obtained through the 
three polarizers. The polarimetric properties of SN\,2014J at different 
late phases will be discussed in a future work. 

After $\sim$600 days past maximum light, the SN became sufficiently dim and 
the count rates at the central pixels of the SN Point Spread Function (PSF) 
became comparable to the bright part of the nebulosity close to the SN. The 
field shows that the SN lies at one end of a dark lane, and just west of a 
bright patch of nebulosity. A background subtraction procedure significantly 
diminishes the time-invariant signals and improves the photometry of evolving 
faint sources. 
{Unfortunately, we found no pre-SN Hubble images, 
either with or without the polarizers, showing the same 
region using filters compatible with our observations.} 
Images obtained on 
March 29 2006 (program $\#$10776; PI:Mountain) with $HST$ ACS/WFC in the 
$F435W$, $F555W$, and $F814W$ were used as background templates for our 
$F475W$, $F606W$, and $F775W$ exposures, respectively. 
For each band, the 
background templates have been scaled and subtracted from the intensity map. 
{The templates have been scaled according to the average flux of four 
local bright sources [(R.A. = 9:55:40.98, Dec. = +69:40:27.16); 
(R.A. = 9:55:41.99, Dec. = +69:40:21.60); 
(R.A. = 9:55:42.84, Dec. = +69:40:31.42); 
(R.A. = 9:55:43.95, Dec. = +69:40:35.47)]. 
} 

{Photometry of SN\,2014J was conducted with a circular aperture of 0.15$\arcsec$ 
(3 pixels in the ACS/WFC FOV) with aperture corrections according to \citet{Hartig_2009} 
and \citet{Sirianni_etal_2005}.} 
The photometry was performed using the {\sc IRAF} \footnote[2]{{\sc iraf} is 
distributed by the National Optical Astronomy Observatories, which is operated by 
the Association of Universities for Research in Astronomy, Inc., under cooperative 
agreement with the National Science Foundation (NSF).}
{\sc apphot} package. 
The residual of the background was 
estimated by the median pixel value of an annulus around the SN. 
Compromising between determining the local background residual with nearby 
pixels and excluding the contamination from resolved interstellar light 
echoes \citep{Yang_etal_2017a}, we choose the inner and outer radii as 
1.2$\arcsec$ (24 pixels) and 1.5$\arcsec$ (30 pixels) for V1 and V2, and 
0.45$\arcsec$ (9 pixels) and 0.75$\arcsec$ (15 pixels) for V3, V4, V5, and V6. 
Table \ref{Table_2} presents the AB magnitudes of SN\,2014J at the six late epochs. 

This photometry strategy has been carried out considering that extremely 
nonuniform background structures dominate the error budget in the late 
phases of the SN\,2014J photometry, especially after V4. For the 
scientific consideration of this study, which is testing the models for 
the light curve evolution at very late phases, the major concern in the 
data reduction procedure is to obtain the correct decline rate of the 
SN light curves. We conducted a sanity check to test the 
reliability of our measurement by performing photometry on differenced 
images from our observations obtained at different epochs. 
{
Observations on V3$\sim$day 649 were subtracted from the observations 
on V4, V5, and V6. 
This directly measures the differential fluxes and therefore the light 
curve decline rate. 
The divergence of magnitude between this estimation and the photometry 
on scaled and background subtracted images are most significant in V6 when the 
SN is faintest, which gives $\sim$0.01, 0.04, and 0.05 magnitude differences in $F475W$, $F606W$, 
and $F775W$, respectively. This difference is $\lesssim$0.01 in V4 and V5. We 
conclude that our photometry is reasonable based on the agreement between these 
two approaches, and the differences represent the systematic uncertainties 
introduced in the use of subtraction templates acquired with different filters. 
The photometric uncertainties we quote include this difference, the Poisson noise of the 
signal, the photon noise of the background, the readout noise contribution 
(3.75 electrons/pixel for ACS/WFC), and the uncertainties in the aperture 
corrections. These quantities were added in quadrature.
} 
The decline rates between all the epochs, calculated from photometry 
shown in Table \ref{Table_3} and measured using this sanity check, agree 
within $\sim$2\% and are smaller than the photometric uncertainties. 

We correct our measurements for both the interstellar dust extinction in 
the SN host galaxy and the Galactic extinction towards SN\,2014J. 
In fact, any imperfection in the extinction correction will only 
affect the individual magnitudes but not the decline rates of the light curves. 
A peculiar extinction law $R_V \sim$1.4 towards the SN\,2014J line of sight 
has been suggested by many studies \citep{Amanullah_etal_2014, Brown_etal_2015, 
Foley_etal_2014, Gao_etal_2015, Goobar_etal_2014}. In this study, we adopt 
$R_V$ = 1.44$\pm$0.03 and $A_V$ = 2.07$\pm$0.18 mag from \citet{Foley_etal_2014} 
for the extinction from the host galaxy and $R_V$ = 3.1 and $E(B-V)=0.054$ 
mag for the Galactic extinction following \citet{Foley_etal_2014} based on 
\citet{Dalcanton_etal_2009} and \citet{Schlafly_etal_2011}. Extinction in the $F475W$, 
$F606W$, and $F775W$ bands has been calculated for each component using a reddening 
law from \citet{Cardelli_etal_1989} with the corresponding $R_V$ value.  
Both components are added to account for the total extinction towards SN\,2014J 
for each {\it HST} ACS bandpass. 

\section{Analysis \label{analysis}}
In this section, we will test different mechanisms powering the late-time 
light curve, and whether the light curve behavior is consistent with the 
prediction for the delayed-detonation and the violent merger 
scenarios following a similar procedure to \citet{Graur_etal_2016} for SN\,2012cg. 
We assume that the ejecta do not interact with any circumstellar material.

\subsection{Pseudo-Bolometric Light Curve \label{analysis_bolo}}
The pseudo-bolometric light curve for SN\,2014J was calculated over a 
wavelength range from 3500$\mathrm{\AA}$ - 9000$\mathrm{\AA}$ based on our multi-band 
optical photometry. We briefly summarize the steps as follows: \\
(1) Based on the lack of significant spectral evolution of SN\,2011fe compared to a spectrum at 
593 days \citep{Graham_etal_2015_11fe}, we assume the MODS/LBT spectrum of 
SN\,2011fe at 1016 days (\citealp{Taubenberger_etal_2015}) represents the major 
spectral features of SN\,2014J on V3$\sim$day 649, V4$\sim$day 796, 
V5$\sim$day 983, and V6$\sim$day 1181. The spectrum was retrieved from the WISeREP archive 
\footnote[3]{
http://wiserep.weizmann.ac.il
}. 
\\
(2) We then perform synthetic photometry on this spectrum for the $F475W$, $F606W$, and $F775W$ bands. \\
(3) We calculate the differences between the synthetic photometry of the SN\,2011fe spectrum 
and our extinction-corrected, observed photometry of SN\,2014J. \\
(4) We calculate the scale factors between the observed and synthetic magnitudes 
in each filter. \\
(5a) We warp the spectrum using a 2$^{\mathrm{nd}}$ order 
polynomial fit to the scale factors determined at the effective wavelength for each filter\footnote[4]{
http://pysynphot.readthedocs.io/en/latest/properties.html\#pysynphot-formula-efflam
}. \\
(5b) Alternatively, for each epoch, we fit a single 
wavelength-independent gray scale across all wavelengths. \\
(6) We iterate steps (2) - (5) until the synthetic and observed photometry
match to better than 0.02 mag in each filter for (5a), or the mean 
difference between the synthetic and the observed photometry converges to its 
minimum value for (5b), for which the standard deviation among the three filters is 0.11 mag. 

\begin{figure}[!htb]
\epsscale{1.0}
\plotone{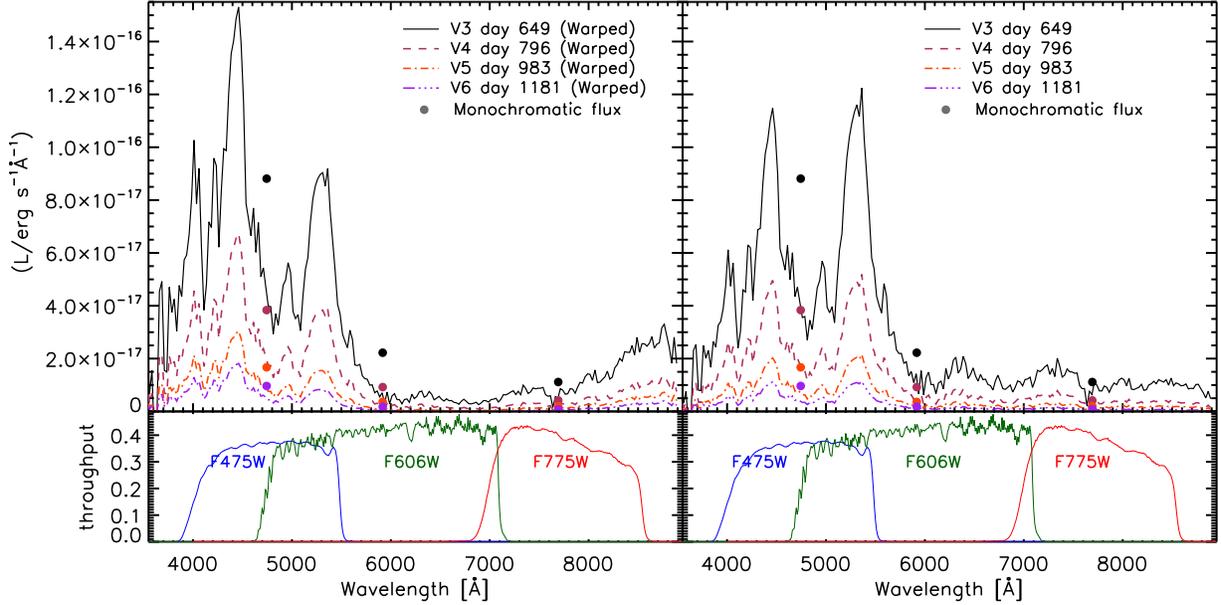}
\caption{\small{The constructed late-time SED for SN\,2014J. Dots show the 
bandpass monochromatic flux from {\it HST} observations at their effective 
wavelengths. {Solid, dashed, dashed-dotted, and triple-dot-dashed lines 
show the spectra constructed with the warping procedure (left panel) and with gray 
scaling (right panel) as described in Section 3, from V3 to V6, respectively.} 
The lower panels present the total bandpass throughput curve ({\it HST} + ACS) 
for our $F475W$, $F606W$, and $F775W$ observations, showing the spectral response 
corresponding to the monochromatic fluxes calculated from the observed photometry.} 
\label{Fig_bolo}} 
\end{figure}

The pseudo-bolometric luminosity for each epoch was obtained by 
integrating the scaled spectrum returned from (5a) or (5b) over the wavelength range 
3500$\mathrm{\AA}$ - 9000$\mathrm{\AA}$. 
The errors on the pseudo-bolometric light curve were computed through 
a Monte Carlo re-sampling approach using the photometric errors.  
The warping in (5a) aims at iteratively producing spectra consistent with the photometry 
which follows a very similar procedure as described in \citet{Shappee_etal_2016}, 
while the scaling in (5b) is less sensitive to the extrapolation of the polynomial correction 
to the spectrum. 

The pseudo-bolometric luminosities calculated from (5a) is on average 13\% higher 
than from (5b). This discrepancy results from the construction of pseudo-bolometric 
light curves. For the scientific consideration of our study, this systematic difference 
does not affect the measurement of the abundance ratio affecting the decline rate of 
the SN luminosity. After correcting this discrepancy, 
the pseudo-bolometric luminosities calculated from these two approaches agree 
within 8\% at all epochs, compatible with the uncertainties of the Monte Carlo approach. 
The error used in fitting the ratio of the isotopes has been estimated by adding this difference 
to the uncertainties obtained from the Monte Carlo approache in quadrature. 
The pseudo-bolometric luminosity of SN\,2014J is listed in Table~\ref{Table_2}. 
The optical pseudo-bolometric luminosity at t$\sim$277 days after the $B$-band maximum 
(log L $\approx 40.28$) is roughly consistent with the UVOIR bolometric luminosity at 
t$\sim$ 269 days (log L $\approx40.35$) estimated from Figure 8 of \citet{Srivastav_etal_2016}. 
Our analysis of the bolometric evolution of SN\,2014J is based on the bolometric luminosity 
obtained with (5b). Qualitatively similar results have been obtained by duplicating the entire 
analysis based on (5a) as follows. 

In Figure~\ref{Fig_bolo} we present the spectra constructed using the warping 
procedure (left panel) and with gray scaling (right panel). For 
comparison, in each upper panel, we overplot the bandpass monochromatic flux 
calculated as the product Total Counts $\times$ PHOTFLAM \footnote[5]{
This can be obtained with the ACS Zeropoints Calculator at 
https://acszeropoints.stsci.edu/}, where PHOTFLAM is the 
inverse sensitivity (in erg cm$^{-2}$ s$^{-1}$ $\rm{\AA^{-1}}$) 
representing a signal of 1 electron per second. 
The lower panels present the total bandpass throughput curve 
({\it HST} + ACS) \footnote[6]{
http://www.stsci.edu/hst/acs/analysis/throughputs} for our $F475W$, $F606W$, 
and $F775W$ observations. The spectra on the left panel are iterated to agree 
quantitatively with the photometry. Visual differences between the monochromatic 
bandpass flux and the spectra arise because PHOTFLAM used for the SED assumes a 
smooth AB spectrum, which is different than the SN spectrum (see 
\citealp{Brown_etal_2016} for a comprehensive discussion).} 

\subsection{Radioactive Decay \label{analysis_decay}}
\begin{figure}[!htb]
\epsscale{1.0}
\plotone{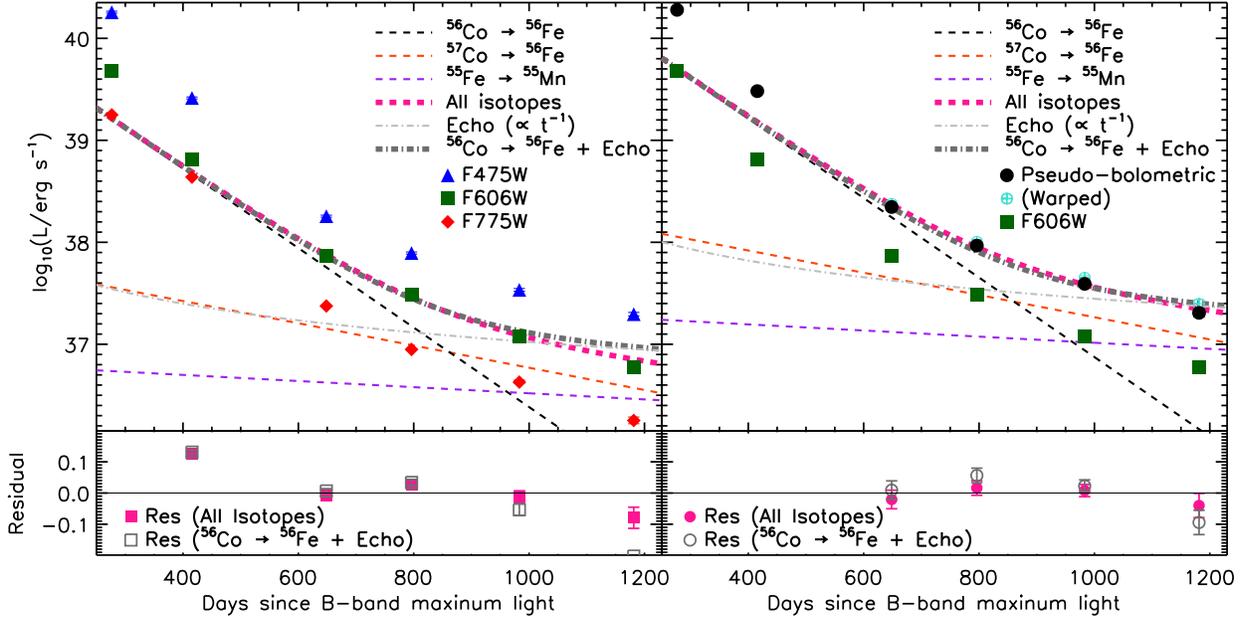}
\caption{\small{Luminosity evolution of the monochromatic fluxes from the broadband 
observations (left panel) and pseudo-bolometric flux (right panel) with 
possible mechanisms explaining the flattening of the light curves of SN\,2014J. 
The left panel presents the 
fitting and residuals of V3 -- V6 based on $F606W$-band observations while the right 
panel shows a similar plot based on the constructed pseudo-bolometric luminosity. 
{In the left panel, we also present the $F475W$ and $F775W$-band observations. The 
$F606W$-band observations together with the pseudo-bolometric light curve constructed 
with warped spectrum (procedure 5a in Section~\ref{analysis}, cyan $\oplus$) 
are shown in the right panel for comparison.} 
The $F606W$-band observations after $\sim$650 days have been assumed to be proportional to 
the bolometric light curves \citep{Milne_etal_2001} and free from possible $\gamma$-ray 
photons. Only observations after 650 days have been fitted with models accounting 
for all the listed isotopes or $^{56}$Co plus a faint, unresolved light echo.  
\label{Fig_fit}}} 
\end{figure}

{In the left panels of Figure~\ref{Fig_fit}, we present the $F475W$, $F606W$, 
and $F775W$-band luminosity of SN\,2014J after correction for the extinction.} 
In addition to fitting the pseudo-bolometric light curve after {$\sim$650 days} 
with the contribution from three decay chains: $^{56}$Co$\rightarrow ^{56}$Fe, 
$^{57}$Co$\rightarrow ^{57}$Fe, and $^{55}$Fe$\rightarrow ^{55}$Mn (an `all 
isotopes' model), we also fit the same model to our $F606W$-band observations.  
Here we have assumed that after $\sim$500 days the $F606W$-band, which is 
centered at wavelength 5888.8$\mathrm{\AA}$ and with a 
width\footnote[7]{where the filter throughput is larger than 0.05\%} of 2570$\AA$, 
captures the dominant Fe features ([Fe II] around 4700$\mathrm{\AA}$ and 5300$\mathrm{\AA}$, 
blended [Fe II]$\lambda$7155 and [Ni II]$\lambda$7378 around 7200$\mathrm{\AA}$; 
\citealp{Taubenberger_etal_2015}) and to be proportional to the bolometric 
light curves as $V$-band observations \citep{Milne_etal_2001}. 

Limited by a small number of visits, we approximate the `all isotopes' 
model with two free parameters: the mass ratio 
$M(^{57}\mathrm{Co})/M(^{56}\mathrm{Co})$, and a scale factor to match 
the $F606W$ photometry (or the pseudo-bolometric luminosity) with the 
model-calculated values. 
Using the solution to the Bateman equation which describes the abundances and 
activities in a decay chain as a function of time (following \citealp{Seitenzahl_etal_2014}), 
and by counting the decay energy carried by charged leptons and X-rays, the luminosity 
contribution from a single decay chain gives:  
\begin{equation}\label{Eqn_1}
L_A (t) = 2.221 \frac{C}{A} \frac{\lambda_A}{\mathrm{day s^{-1}}} \frac{M(A)}{M_{\odot}}
\frac{q^l_A + q^X_A}{\mathrm{keV}} \mathrm{exp} (-\lambda_A t_e) \times 10^{43} \mathrm{erg s^{-1}}
\end{equation}
where $C$ is a scaling factor, 
$A$ gives the corresponding atomic number, 
$\lambda_A$ is the inverse mean lifetime 
($\lambda_A = \tau_A^{-1} = \mathrm{ln(2)}/t_{1/2, A}$), 
$M(A)$ is the total mass of a certain decaying element, 
$q^l_A$ and $q^X_A$ are the average energies per decay carried by charged 
leptons and X-rays, respectively, 
and $t_e$ is the time since explosion. 
Due to the limited data points in our late-time photometry, we used 
a ratio of $M(^{57}Co)/M(^{55}Fe) \approx 0.8$ (model rpc32; 
\citealp{Ohlmann_etal_2014}). The values of $\lambda_A$, $q^l_A$ and $q^X_A$ 
used here are obtained from Table 1 of \citet{Seitenzahl_etal_2009} and 
Table 2 of \citet{Seitenzahl_etal_2014}. 
We justify our assumptions as follows: (1) The total deposition function 
is determined by both the net deposition functions for $\gamma-$rays and positrons. 
The $\gamma-$rays produced by the annihilation of the positrons are subject to both 
deposition functions. By simply assuming the radioactive source is confined to the 
center of a spherical distribution of ejecta yields a fraction $1 - e^{-\tau_{\gamma}}$ 
of the energy produced by $\gamma-$rays would be left behind in the 
ejecta \citep{Swartz_etal_1991}. The $\gamma-$ray optical depth $\tau_{\gamma}$ drops 
significantly as $t^{-2}$ and we neglect contributions from $\gamma$-rays 
because the SN ejecta became transparent to $\gamma$-rays at $t\gtrsim$500 days 
\citep{Milne_etal_2001}; (2) {Limited by a small number of photometric points}, 
we begin by fitting Equation \ref{Eqn_1} assuming full trapping of positrons/electrons. 
In other words, we assume positrons, electrons, and X-rays are fully trapped, instantaneously 
deposited, and radiate their energy. One should also note that very recently, 
\citet{Dimitriadis_etal_2017} found that the late-time bolometric light curve 
of SN\,2011fe is consistent with both models{: either a model that allows for 
positron/electron escape, or a model that has complete positron/electron trapping 
but do allow for redistribution of flux to the mid-far IR.} 

The luminosity contribution from each decay channel is shown in 
Figure~\ref{Fig_fit}. The total luminosity given by these decay chains is 
represented by the pink dashed line. In the left panel, 
{we show that a 
mass ratio of $M(^{57}\mathrm{Co})/M(^{56}\mathrm{Co}) = 0.065_{-0.004}^{+0.005}$ 
gives the best fit to the `all isotopes' model based on the $F606W$-band 
observations after $t\sim500$ days (V3 -- V6).} 
The dot-dashed gray 
lines show the model including the luminosity from $^{56}$Co decay and 
possible reflections from an unresolved $t^{-1}$ light echo (see 
\citealp{Graur_etal_2016}). 
In the right panel, we show the same trend in a similar fitting based on the 
pseudo-bolometric light curve, 
{which the mass ratio gives 
$M(^{57}\mathrm{Co})/M(^{56}\mathrm{Co}) = 0.066_{-0.008}^{+0.009}$. 
We also tested the same abundance ratio, using a fit based on the 
pseudo-bolometric light curve constructed with the warped spectrum (procedure 
5a in Section~\ref{analysis}). A similar mass ratio of 
$M(^{57}\mathrm{Co})/M(^{56}\mathrm{Co}) = 0.078_{-0.010}^{+0.011}$ has been obtained. 
}

\subsection{Light Echoes? \label{analysis_echo}}
If light echoes dominate the late time signal from the SN, we may expect 
a significant profile change or centroid drift if the circumstellar 
matter is distributed at sufficiently large distances from the SN. Light 
scattered by dust at such distances can produce measurable distortions to 
the image profiles if the scattered light dominates the total observed 
flux. At the distance of SN\,2014J, 1 light year corresponds to 0.17 
{\it HST} ACS/WFC pixels. Depending on the dust distribution, we may 
expect the stellar profiles to become non-point like, or the centroid of 
the stellar profile to drift at late times. We have checked the stellar 
profiles and found no significant deviations from a point source at all 
epochs of our observations. In the following, we provide a comprehensive 
check on the centroid position of the SN.

\begin{figure}[!htb]
\epsscale{1.0}
\plotone{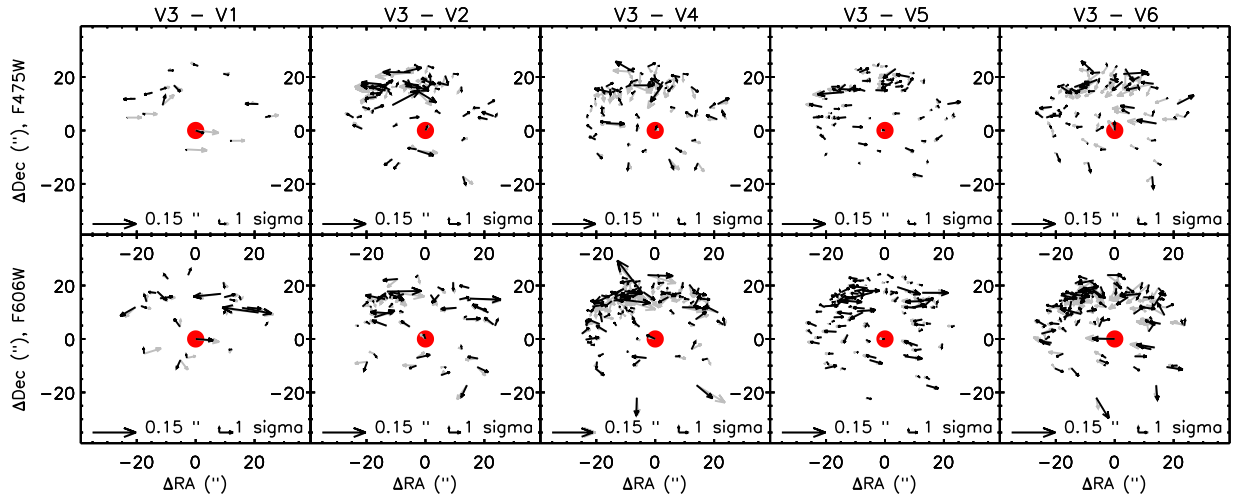}
\caption{\small
Astrometric comparisons of different visits. The $x$ and $y$-axis are the position 
of bright sources relative to SN 2014J, most of which are HII regions in M82. 
{The SN is represented by the red dot at the origin.} 
The gray and black arrows are the relative motion between different visits 
prior to and after a linear regression with the $RA$ and $Dec$. 
A 1-$\sigma$ displacement calculated based on all the presented sources 
and scales are provided at the bottom right of each panel. No significant 
positional drift of the SN is found among all the cases, suggesting the 
absence of any circumstellar light echoes around 1 light year from the SN on 
the plane of the sky. 
\label{Fig_center}} 
\end{figure}

The barycenter of the stars and HII regions around SN\,2014J were measured to 
estimate a possible change in the relative position of the light emission of the 
SN. The precision is limited by the scarcity of stars in the immediate vicinity 
of the SN, as well as the uncharacterized field distortions caused by 
ACS/WFC polarizers (see, i.e., Section 5.3 of \citealp{Gonzaga_etal_2012}). 
Figure~\ref{Fig_center} presents the apparent shift in position measured from our 
observations in $F475W$ and $F606W$. The $RA$ and $Dec$ were calculated 
using the image from V3, with the SN at the origin of the coordinates. 
The gray arrows show the vector difference of the originally measured positions 
of the source on two different epochs. The black arrow shows the same vector 
after a 2-D linear regression to remove the dependence on $RA$ and $Dec$, 
which may be caused by residual errors of astrometric calibrations. The 
linear regression was found to be able to reduce the shift 
significantly in all cases. The reference objects for astrometric 
comparisons were selected within a radius of 500 pixels of the position of 
the SN. The FWHM of the objects was restricted to be less than 8 pixels. 
Only a small number objects in the earliest epoch V1 satisfy these criteria due 
to the relatively short exposure time. 

In Figure~\ref{Fig_center}, the upper panels present the measurements based on 
the highest S/N $F475W$-band exposures, and the lower panels present the same 
figures for $F606W$. 
{For V5 and V6 when the the SN became sufficiently dim, 
to minimize the effect of local background, the centroid of the SN was determined 
based on scaled and background subtracted images. 
For instance, in the upper row, the first panel presents the comparison 
between V3 and V1. The SN (red dot) exhibits an apparent motion of 0.079$\arcsec$ (gray arrow); 
after linear regression with the $RA$ and $Dec$, this reduces to 0.029$\arcsec$ (black arrow). 
This is in agreement with all the other objects in the field, which show an 
average distance shift of 0.022$\arcsec$ and an RMS of 0.014$\arcsec$. The 
second panel presents the comparison between V3 and V2. The SN exhibits an 
apparent drift in position of 0.020$\arcsec$; after linear regression this 
reduces to 0.016$\arcsec$. The field objects exhibit an average drift of 0.036$\arcsec$ 
and an RMS of 0.023$\arcsec$. This implies that the 
position drift of the SN is significantly lower than the average of the field objects. The 
third to the fifth panels present the comparison between V3 and V4, V3 and V5, V3 and V6, respectively. 
After linear regression with $RA$ and $Dec$, 
using the stars around the SN, the drift of the SN compared to the average drift$\pm$RMS gives: 
0.015$\arcsec$ vs. 0.024$\pm$0.016$\arcsec$, 
0.008$\arcsec$ vs. 0.021$\pm$0.016$\arcsec$, and 
0.031$\arcsec$ vs. 0.027$\pm$0.017$\arcsec$, respectively. 
An upper bound on the centroid position drift of the SN between V3 
and another epoch is thus observed to be the sum of the SN drift and the RMS of the drift 
measured from field objects. In each of these cases, this upper bound has found to be larger 
than the average drift of the field objects, which implies that there is no apparent 
position drift of the SN. 
Similar results were obtained for 
$F606W$-band exposures. In all cases, we have not observed a significant position 
drift of the SN. The only exception is the 0.077$\arcsec$ vs. 0.031$\pm$0.018$\arcsec$ in 
V3 compared to V6, $F606W$. Considering no drift was found in the same epoch of $F475W$ 
and the low signal-to-noise ratio of the $F606W$ observation, we do not consider significant 
drift of the SN in V6. 
The absence of such drift sets a strong constraint on the nature 
of the late time emission from SN 2014J. 
If the significant flattening in $F606W$-band and pseudo-bolometric light curves 
is due to light echoes, the dust must be lie within 0.017$\arcsec$ of the SN. 
}

\begin{figure}[!htb]
\epsscale{0.7}
\plotone{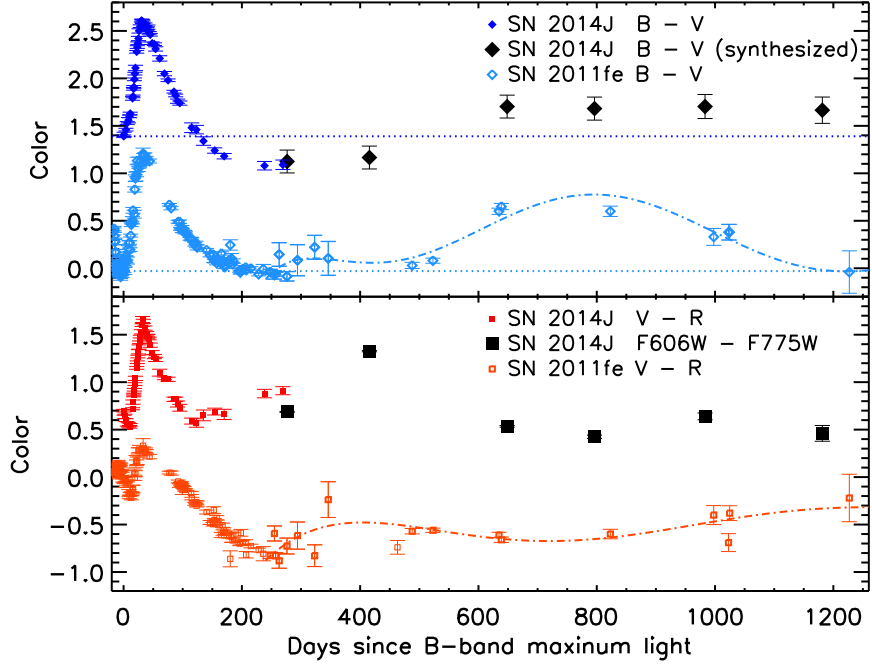}
\caption{\small{Comparison of the color evolution of SN\,2014J and SN\,2011fe until very late 
phases to address the possibility of an unresolved light echo within the PSF. The top 
panel presents the $B-V$ color calculated with {\sc pysynphot} based on the gray-scaled 
spectrum of SN\,2014J at late epochs and the $B-V$ color of SN\,2014J from t$\sim$-8 to 
269 days \citet{Srivastav_etal_2016}. The $B-V$ color curve of SN\,2011fe at early 
\citep{Zhang_etal_2016} and at late \citep{Shappee_etal_2016} epochs are shown for comparison. 
Dotted-dashed lines show polynomial fit to the color evolution after day $\sim$250 and 
horizontal dashed lines indicate the color at the SN maximum. The fact that SN\,2014J 
has become redder than it was at peak and SN\,2011fe at similar epochs limits the flux any 
light echo could be contributing. The bottom panel gives the evolution of the $F606W-F775W$  
color of SN\,2014J and the $V-R$ color of SN\,2011fe for comparison.} 
\label{Fig_color_test}} 
\end{figure}

{Here we address the possibility of an unresolved light echo within the PSF 
of the late-time source at the SN position. 
Our photometry allows us to measure the $F475W-F606W$ and $F606W-F775W$ colors 
at very late phases. We also compared the late-time color evolution of SN\,2014J 
with SN\,2011fe, which does not exhibit evident flux contribution from the light 
echoes. Light-echo flux is dominated by the light of the SN around its peak, and 
scattering by dust favors blue light. At extremely late phases, 
when light from the SN may no longer dominate over the scattered light echoes, 
the color of the integrated flux can appear to be bluer by a few tenths of a magnitude 
\citep{Rest_etal_2012_color, Graur_etal_2016}. A redder color measured 
at very late time, therefore, would suggest the absence of a light echo. 
In Figure~\ref{Fig_color_test}, we present the comparison of the late-time 
color evolution of SN\,2014J and SN\,2011fe. 
The $B$ and $V$-band AB magnitudes of SN\,2014J were calculated with {\sc pysynphot} 
using the gray-scaled spectrum introduced in (5b) in Section~\ref{analysis}. 
Systematical differences between the synthetic photometry in $F475W$ and $F606W$ 
on the gray-scaled and the {\it HST} photometry have been included when calculating 
the error in the $B-V$ color of SN\,2014J. 

The $B-V$ color of SN\,2014J from t$\sim$-8 to 269 days has been calculated based 
on the photometry of \citet{Srivastav_etal_2016}. The $B-V$ color curve of SN\,2011fe 
at early \citep{Zhang_etal_2016} and at late \citep{Shappee_etal_2016} phases are 
shown for comparison. We note that from day $\sim$140 to 500, the $B-V$ color of 
SN\,2014J appears to be bluer than it was around the maximum light (see, i.e., 
Figure~\ref{Fig_color_test}). A similar effect can be expeceted if SN\,2014J was 
contaminated by light echoes. The color of SN\,2014J at day $\sim$650 to 
1200 is however redder in $B-V$ by $\sim$0.3 magnitude, and the color-evolution of SN\,2014J 
also shows a similar trend to that of SN\,2011fe at the same phase. 
Spectra of SN\,2011fe at day $\sim$1000 detected no trace of a light echo 
\citep{Graham_etal_2015_11fe, Taubenberger_etal_2015}. Thus, the similarity in 
the late-time color evolutions of SN\,2014J and SN\,2011fe, together with our 
astrometric analysis, lead us to argue that the luminosity measurement of SN\,2014J 
was not contaminated by a light echo at day $\sim$650 to 1200. 
}

\section{Discussion and Summary}
Table \ref{Table_3} shows the decline rate of the light curves at 
different epochs. Before $t\sim$600 days, the SN dims more rapidly than 
the light curve powered solely by the $^{56}$Co decay. The $\gamma$-ray 
energy deposition becomes no longer significant after $\sim$200 days, 
therefore, a substantial fraction of the flux may be shifting out of the optical 
bands into the infrared. Similar behavior has been discussed in the case 
of SN\,2011fe \citep{Kerzendorf_etal_2014} and SN\,2003hv 
\citep{Leloudas_etal_2009}. After $t\sim$600 days, a slower decay 
can be identified in all the $F475W$, $F606W$, and $F775W$-bandpasses. 

{Some observations of nearby type Ia SNe show that their bolometric 
light curves at late phases follow the $^{56}$Co decay channel 
(\citealp{Cappellaro_etal_1997, Sollerman_etal_2004, Lair_etal_2006, 
Stritzinger_etal_2007, Leloudas_etal_2009}). These observations 
suggest that a turbulent, confining 
magnetic field traps the positrons, resulting in local energy deposition (see
\citealp{Chan_etal_1993, Milne_etal_1999, Milne_etal_2001, Penney_etal_2014}). 
In contrast, $^{56}$Co positron escape has been suggested in some cases 
\citep{Milne_etal_1999, Milne_etal_2001}. 
As the ejecta expand over time, the pre-configured magnetic field weakens 
to the point that the Larmor radius exceeds the size of the turbulence
(see \citealp{Penney_etal_2014}). 
} 

{
The late-time pseudo-bolometric decline rate of SN\,2014J during 
day 277 to day 416 (1.432$\pm$0.044 mag per 100 days) and day 416 to 
day 649 (1.219$\pm$0.038) is larger than the predicted decay rate of radioactive 
$^{56}$Co (0.98 mag per 100 days). This may be caused by the positron escape 
which would produce a faster decay rate. A similar decline rate can also be 
seen in the quasi-bolometric light curve of SN\,2014J at $\sim$day 238 to 269 
(i.e., $\sim$1.3 mag per 100 days, \citealp{Srivastav_etal_2016}). 
Qualitatively speaking, at these intermediate epochs, the contributions from 
$\gamma$-rays may still be non-negligible since the SN ejecta may not have 
become transparent to $\gamma$-ray photons. 
} 

We fit both the $F606W$-band and a `pseudo-bolometric' light curve 
using Bateman's equation for the luminosity contribution of the $^{56}$Co, 
$^{57}$Co, and $^{55}$Fe decay channels. The best fit to the 
pseudo-bolometric light curve and the $F606W$-band light curve give a 
mass ratio $M(^{57}\mathrm{Co})/M(^{56}\mathrm{Co}) = 0.065_{-0.004}^{+0.005}$ 
and 0.066$_{-0.008}^{+0.009}$, respectively. Assuming the same mass ratio yields for 
isotopes of the same iron-group elements (see \citealp{Graur_etal_2016}, 
based on \citealp{Truran_etal_1967} and \citealp{Woosley_etal_1973}), 
our measurements correspond to $\sim$3 times the 
$M(^{57}\mathrm{Fe})/M(^{56}\mathrm{Fe})$ ratio of the Sun (i.e., 
$\sim$0.0217, see Table 3 of \citealp{Asplund_etal_2009}). 
This is higher than the solar ratio $\sim$1.8 predicted for the W7 model (calculated from 
Table 3 of \citealp{Iwamoto_etal_1999}), 
{and the solar ratio $\sim$1.7 predicted for the near-Chandrasekhar-mass three-dimensional 
delayed-detonation model N1600 (calculated from Table 2 of \citealp{Seitenzahl_etal_2013})}. 
The $M(^{57}\mathrm{Fe})/M(^{56}\mathrm{Fe})$ ratio in our measurements is also 
higher compare to the ratios $\sim$2 and $\sim$1.1 
suggested by the late-time quasi-bolometric light curve analysis on 
SN\,2012cg \citep{Graur_etal_2016} and SN\,2011fe \citep{Shappee_etal_2016}. 
A higher metallicity progenitor could decrease the production of 
$^{56}$Ni and result in a higher $M(^{57}\mathrm{Ni})/M(^{56}\mathrm{Ni})$ 
ratio \citep{Seitenzahl_etal_2013}. An enhancement of neutron 
excess due to electron captures in the deflagration wave could lead to 
the same effect. 

{It has been suggested that beyond $\sim$500 days in the ejecta, energy is 
shifted from the optical and near-infrared to the mid- and far-infrared 
(referred as the infrared catastrophe, \citealp{Axelrod_1980}, and 
see \citealp{Fransson_etal_1996, Fransson_etal_2015}). 
The $V$ or optical luminosity may not represent the 
actual behavior of the bolometric light curves. This has never been 
observed so far in any type Ia SNe (e.g., \citealp{Sollerman_etal_2004, 
Leloudas_etal_2009, McCully_etal_2014, Kerzendorf_etal_2014, Graur_etal_2016, 
Shappee_etal_2016}). However, \citet{Dimitriadis_etal_2017} suggested that 
the evolution of SN\,2011fe, around 550 to 650 days, is consistent with both 
a model that allows for {positron/electron} escape and a model allowing 
for a redistribution of flux from optical to the mid-far infrared. In our study, 
we fitted the $F606W$-band and optical bolometric luminosity after $\sim$650 
days and do not consider the infrared catastrophe. Future studies based on a 
larger sample will be able to help distinguish these two possible scenarios. 
} 

{As suggested by \citet{Kerzendorf_etal_2017_11fe}, although the flattening 
of the late-time light curves of SN\,2014J can be well-explained by additional 
energy input from the decay of $^{57}$Co, we concede that one cannot draw 
strong conclusions from the current observation due to the uncertain physical 
processes. The determination of a precise isotopic abundance does require 
detailed modeling of the processes. 
Another mechanism that may plausibly explain the late-time luminosity flattening 
is the survival of the donor WD after the explosion. A small amount of 
$^{56}$Ni-rich material synthesized by the primary WD's explosion at low 
velocities might remain gravitationally bound and captured by the surviving WD companion 
\citep{Shen_etal_2017_survive}. The lack of electrons on the surface of the donor 
WD significantly reduces the decay rates of $^{56}$Ni and $^{56}$Co than electron 
capture \citep{Sur_etal_1990, daCruz_etal_1992}. The radioactive decay is delayed 
and thus the surviving WD can be another source of late-time type Ia SN luminosity. 
Future observations of type Ia SNe at extremely late phases will be important 
to understanding the physical processes at this late stage 
and further testing the explosion mechanisms of type Ia SNe.
} 

In summary, our multi-band photometry of SN\,2014J out to 1181 days past the 
$B-$band maximum light clearly detected the flattening due to extra luminosity 
contributions other than the decay of $^{56}$Co. {We conclude that the high 
$M(^{57}\mathrm{Ni})/M(^{56}\mathrm{Ni})$ ratio estimated from the late-time 
luminosity evolution of SN\,2014J favors a near-Chandrasekhar mass explosion 
model such as W7 of \citet{Iwamoto_etal_1999}.} 
Any significant circumstellar light echoes beyond 0.3 pc on the plane of 
the sky can be excluded by our astrometric analysis. The observations strongly 
suggest additional heating from internal conversion and Auger electrons of 
$^{57}$Co$\rightarrow ^{57}$Fe; however, one should be cautious on the high 
mass ratio of $^{57}$Ni to $^{56}$Ni. Systematical uncertainties from the SED 
construction procedure, especially the missing information from NIR 
observations and the interpolation of the SED based on limited bandpass 
coverage should not be ignored (i.e., see \citealp{Brown_etal_2016}). 
Additionally, the reliability of approximating the bolometric luminosity 
evolution after $t\sim$650 days with the $F606W$-band emission requires more 
careful justification. 

\acknowledgments 
The authors are grateful to Dave Borncamp and the {\it HST} ACS team in fixing the 
distortion correction issues in ACS/WFC polarized images. 
{We would like to thank the anonymous referee and Or Graur for very helpful 
discussion and constructive suggestions that improved the paper.} 
Some of the data used in this study were obtained from the Mikulski 
Archive for Space Telescopes (MAST). STScI is operated by the Association 
of Universities for Research in Astronomy, Inc., under NASA contract NAS5-26555. 
Support for MAST for non-HST data is provided by the NASA Office of Space Science 
via grant NNX09AF08G and by other grants and contracts.
This work also made use of the Weizmann interactive supernova data
repository (WISeREP). 
The supernova research by Y. Yang,
P. J. Brown, and L. Wang is supported by NSF grant AST-0708873.
P. J. Brown was partially supported by a Mitchell Postdoctoral
Fellowship.  Y. Yang and M. Cracraft also acknowledge support from
NASA/STScI through grant HST-GO-13717.001-A, grant HST-GO-13717.001-A, 
HST-GO-14139.001-A, and HST-GO-14663.001-A. 
The research of Y. Yang 
is supported through a Benoziyo Prize Postdoctoral Fellowship.
The research of J. Maund 
is supported through a Royal Society University Research Fellowship.
L. Wang is supported by
the Strategic Priority Research Program ``The Emergence of Cosmological
Structures'' of the Chinese Academy of Sciences, Grant No. XDB09000000.
L. Wang and X. Wang are supported by the Major State Basic Research
Development Program (2013CB834903), and X. Wang is also supported by
the National Natural Science Foundation of China (NSFC grants 11178003
and 11325313).

\begin{deluxetable}{ccccccccccc}
\tablewidth{0pc}
\tabletypesize{\scriptsize}
\tablecaption{Log of Observations of SN\,2014J with $HST$ ACS/WFC POLV \label{Table_1}}
\tablehead{
\colhead{Filter} \vspace{-0.0cm}  & \colhead{Polarizer}  & \colhead{Date}  & \colhead{Exp}  & \colhead{Phase$^a$} & \colhead{Date} & \colhead{Exp} & \colhead{Phase$^a$} & \colhead{Date} & \colhead{Exp} & \colhead{Phase$^a$} \\
\colhead{} &  &  \colhead{(UT)} & \colhead{(s)}  & \colhead{(Days)} & \colhead{(UT)} & \colhead{(s)} & \colhead{(Days)} & \colhead{(UT)} & \colhead{(s)} & \colhead{(Days)}  }
\startdata
F475W & POL0V   & 2014-11-06 &  3$\times$130  & 276.5   & 2015-03-25 &  3$\times$400  & 415.6  & 2015-11-12 &  4$\times$1040 & 648.5  \\
F475W & POL120V & 2014-11-06 &  3$\times$130  & 276.5   & 2015-03-25 &  3$\times$400  & 415.6  & 2015-11-12 &  4$\times$1040 & 648.7  \\
F475W & POL60V  & 2014-11-06 &  3$\times$130  & 276.5   & 2015-03-25 &  3$\times$400  & 415.7  & 2015-11-12 &  4$\times$1040 & 648.8  \\
F606W & POL0V   & 2014-11-06 &  2$\times$40   & 276.6   & 2015-03-27 &  3$\times$60   & 417.9  & 2015-11-12 &  4$\times$311  & 649.0  \\
F606W & POL120V & 2014-11-06 &  2$\times$40   & 276.6   & 2015-03-27 &  3$\times$60   & 418.0  & 2015-11-13 &  4$\times$311  & 649.0  \\
F606W & POL60V  & 2014-11-06 &  2$\times$40   & 276.6   & 2015-03-27 &  3$\times$60   & 418.0  & 2015-11-13 &  4$\times$311  & 649.1  \\
F775W & POL0V   & 2014-11-06 &  2$\times$30   & 276.6   & 2015-03-27 &  3$\times$20   & 418.0  & 2015-11-12 &  4$\times$100  & 648.5  \\
F775W & POL120V & 2014-11-06 &  1$\times$55   & 276.6   & 2015-03-27 &  3$\times$20   & 418.0  & 2015-11-12 &  4$\times$100  & 648.7  \\
F775W & POL60V  & 2014-11-06 &  1$\times$55   & 276.6   & 2015-03-27 &  3$\times$20   & 418.0  & 2015-11-12 &  4$\times$100  & 648.9  \\
                \hline
F475W & POL0V   & 2016-04-08 &  4$\times$1040 & 796.2   & 2016-10-12 &  4$\times$1040 & 983.1  & 2017-04-28 &  4$\times$1040 & 1181.3 \\
F475W & POL120V & 2016-04-08 &  4$\times$1040 & 796.4   & 2016-10-12 &  4$\times$1040 & 983.3  & 2017-04-28 &  4$\times$1040 & 1181.4 \\
F475W & POL60V  & 2016-04-08 &  4$\times$1040 & 796.6   & 2016-10-12 &  4$\times$1040 & 983.4  & 2017-04-28 &  4$\times$1040 & 1181.5 \\
F606W & POL0V   & 2016-04-08 &  4$\times$311  & 796.8   & 2016-10-14 &  3$\times$360  & 985.1  & 2017-04-28 &  3$\times$360  & 1181.7 \\
F606W & POL120V & 2016-04-08 &  4$\times$311  & 796.8   & 2016-10-14 &  3$\times$360  & 985.1  & 2017-04-28 &  3$\times$360  & 1181.7 \\
F606W & POL60V  & 2016-04-08 &  4$\times$311  & 796.9   & 2016-10-14 &  3$\times$360  & 985.1  & 2017-04-28 &  3$\times$360  & 1181.7 \\
F775W & POL0V   & 2016-04-08 &  4$\times$100  & 796.2   & 2016-10-12 &  4$\times$202  & 983.1  & 2017-04-28 &  4$\times$202  & 1181.3 \\
F775W & POL120V & 2016-04-08 &  4$\times$100  & 796.4   & 2016-10-12 &  4$\times$202  & 983.3  & 2017-04-28 &  4$\times$202  & 1181.4 \\
F775W & POL60V  & 2016-04-08 &  4$\times$100  & 796.6   & 2016-10-12 &  4$\times$202  & 983.4  & 2017-04-28 &  4$\times$202  & 1181.5 \\
\enddata
\tablenotetext{a}{Days since B maximum on 2014 Feb. 2.0 (JD 245 6690.5).}
\label{Table_1}
\end{deluxetable}

\begin{table}
\caption{{\it HST} ACS/WFC late-time Photometry of SN\,2014J}
\begin{scriptsize}
\begin{tabular}{c|cc|cc|cc|c}
\hline
Filter &  \multicolumn{2}{c|}{$F475W$}   &  \multicolumn{2}{c|}{$F606W$}    &  \multicolumn{2}{c}{$F775W$}     &  log L$^b$\\
Visit  &  Phase$^a$ &  AB Magnitude      &  Phase$^a$  &  AB Magnitude      &  Phase$^a$  &  AB Magnitude      & (erg s$^{-1}$)\\
\hline
1      &  276.5     &  17.363$\pm$0.003  &  276.6      &  17.429$\pm$0.003  &  276.6      &  16.742$\pm$0.004  & 40.279$\pm$0.017 \\
2      &  415.6     &  19.464$\pm$0.003  &  418.0      &  19.602$\pm$0.004  &  418.0      &  18.276$\pm$0.005  & 39.482$\pm$0.018 \\
3      &  648.7     &  22.363$\pm$0.004  &  649.0      &  21.962$\pm$0.005  &  648.7      &  21.427$\pm$0.007  & 38.346$\pm$0.030 \\
4      &  796.4     &  23.266$\pm$0.007  &  796.8      &  22.917$\pm$0.013  &  796.4      &  22.492$\pm$0.012  & 37.968$\pm$0.023 \\
5      &  983.3     &  24.169$\pm$0.016  &  985.1      &  23.936$\pm$0.032  &  983.3      &  23.294$\pm$0.016  & 37.592$\pm$0.019 \\
6      & 1181.4     &  24.765$\pm$0.026  & 1181.7      &  24.695$\pm$0.060  & 1181.4      &  24.234$\pm$0.057  & 37.308$\pm$0.039 \\
\hline
\end{tabular}
\label{Table_2}
\\
{$^a$}{Approximate days after $B$ maximum, 2014 Feb. 2.0 (JD 245 6690.5). } \\
{$^b$}{Phases in $F475W$ have been used. }
\end{scriptsize}
\end{table}

\begin{deluxetable}{ccccc}
\tablewidth{0pc}
\tabletypesize{\scriptsize}
\tablecaption{{\it HST} Late-time light curve decline rate of SN\,2014J \label{Table_3}}
\tablehead{
\colhead{Period$^a \setminus$Filter} \vspace{-0.0cm}  & \colhead{$F475W$}  & \colhead{$F606W$} & \colhead{$F775W$} & \colhead{Pseudo-bolometric} \\
\colhead{(Days)} & \colhead{($\Delta$mag/100 days)}  & \colhead{($\Delta$mag/100 days)} & \colhead{($\Delta$mag/100 days)}  & \colhead{($\Delta$mag/100 days)} }
\startdata
277 -- 416  &   1.511$\pm$0.003  &  1.532$\pm$0.004 &  1.079$\pm$0.004  &  1.432$\pm$0.044 \\
416 -- 649  &   1.245$\pm$0.002  &  1.024$\pm$0.003 &  1.370$\pm$0.003  &  1.219$\pm$0.038 \\
649 -- 796  &   0.611$\pm$0.006  &  0.646$\pm$0.009 &  0.721$\pm$0.009  &  0.640$\pm$0.064 \\
796 -- 983  &   0.483$\pm$0.009  &  0.540$\pm$0.018 &  0.429$\pm$0.011  &  0.503$\pm$0.040 \\
983 -- 1181 &   0.301$\pm$0.015  &  0.387$\pm$0.035 &  0.474$\pm$0.030  &  0.358$\pm$0.055 \\
\enddata
\tablenotetext{a}{Approximate days after $B$ maximum, 2014 Feb. 2.0 (JD 245 6690.5). }
\end{deluxetable}

\clearpage


\end{document}